\documentclass{article}

\usepackage{arxiv}

\usepackage[utf8]{inputenc} 
\usepackage[T1]{fontenc}    
\usepackage{hyperref}       
\usepackage{url}            
\usepackage{booktabs}       
\usepackage{amssymb,amsmath,amsfonts,amsthm}       
\usepackage{nicefrac}       
\usepackage{microtype}      
\usepackage{graphicx,import,rotating}
\usepackage{doi}

\usepackage{caption,subcaption}
\usepackage{tabularx,multirow}

\usepackage{xcolor}
\usepackage{mathrsfs}
\usepackage{appendix,textcomp}
\usepackage{algorithm,algorithmicx,algpseudocode,program,listings}

\algblockdefx[LBM]{StartLBM}{EndLBM}[1][SolidLBM]{\textbf{Start}\ #1}{\textbf{End}}
\algblockdefx[Mesher]{StartMesher}{EndMesher}[1][Preprocesser]{\textbf{Start}\ #1}{\textbf{End}}
\algblockdefx[Solver]{StartSolver}{EndSolver}[1][Solver]{\textbf{Start}\ #1}{\textbf{End}}

\renewcommand{\vec}{\boldsymbol}        

\newcommand{\argum}{(\vec{x},t)}
\newcommand{\abs}[1]{\ensuremath{\left\vert#1\right\vert}}

\title{Dynamic Propagation of Mode III Cracks in a Lattice Boltzmann Method for Solids}

\author{\href{https://orcid.org/0000-0003-4819-2198}{\includegraphics[scale=0.06]{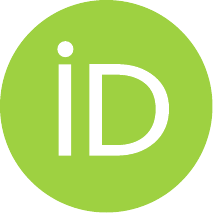}\hspace{1mm}Henning M\"uller}${}^{1}$\thanks{E-mail address for correspondence: \texttt{henning.mueller@tu-darmstadt.de}}, %
    Ali Touil${}^{2}$, %
    \href{https://orcid.org/0000-0002-0839-6519}{\includegraphics[scale=0.06]{orcid.png}\hspace{1mm}Alexander Schl\"uter}${}^{2}$, %
    Ralf M\"uller${}^{1}$\\[8pt]
    ${}^{1}$~Institute for Mechanics, Continuum Mechanics Group, Technische Universit\"at Darmstadt, Germany\\
    ${}^{2}$~Institute of Applied Mechanics, Technische Universit\"at Kaiserslautern, Germany
}

\date{}

\renewcommand{\headeright}{}
\renewcommand{\undertitle}{}
\renewcommand{\shorttitle}{Dynamic Cracks in an LBM for solids}

\hypersetup{
pdftitle={Dynamic Propagation of Mode III Cracks in a Lattice Boltzmann Method for Solids},
pdfsubject={physics.comp-ph, cs.CE},
pdfauthor={Henning~M\"uller, Ali~Touil, Alexander~Schl\"uter, Ralf~M\"uller},
pdfkeywords={Lattice Boltzmann Method, solids, dynamic fracture mechanics, computational solid mechanics},
}

\begin{document}
\maketitle

\begin{abstract}
        This work presents concepts and algorithms for the simulation of dynamic fractures with a Lattice Boltzmann method (LBM) for linear elastic solids.
        This LBM has been presented previously and solves the wave equation, which is interpreted as the governing equation for antiplane shear deformation.
        Besides the steady growth of a crack at a prescribed crack velocity, a fracture criterion based on stress intensity factors (SIF) has been implemented.
        This is the first time, that crack propagation with a mechanically relevant criterion is regarded in the context of LBMs.
        Numerical results are examined to validate the proposed method.

        The concepts of crack propagation introduced here are not limited to mode III cracks or the simplified deformation assumption of antiplane shear.
        By introducing a rather simple processing step into the existing LBM at the level of individual lattice sites, the overall performance of the LBM is maintained.

        Our findings  underline the validity of the LBM as a numerical tool to simulate  solids in general as well as dynamic fractures in particular.
\end{abstract}

\keywords{Lattice Boltzmann Method \and solids \and  dynamic fracture mechanics \and  computational solid mechanics}

\section{Introduction}
\label{sec:intro}

Regarding solid bodies and structures, not only the deformation under external loads is of interest. Different fields of science and engineering, such as geophysics or civil engineering, are also concerned with the mechanism of fracture and the study thereof.
Thus, a number of different numerical techniques for the simulation of dynamic crack propagation and other fracture related phenomena have emerged. More prominent among these are the finite element method (FEM)~\cite{barsoum_use_1976,moes_finite_1999} the boundary element method~\cite{erdogan_numerical_1973,zhang_new_1989}, or more recently peridynamics~\cite{silling_meshfree_2005,ha_characteristics_2011} and phase field methods~\cite{bourdin_variational_2008,schluter_phase_2014}.

Lattice Boltzmann methods (LBM)~\cite{kruger_lattice_2017} are another approach to simulations in engineering and are now widely used in computational fluid dynamics. The principle, however, can be adapted to different problems in various disciplines of science~\cite{succi_lattice_2018,solorzano_lattice_2018}.
The usage of LBM in computational solid mechanics is recently developed~\cite{marconi_lattice_2003,obrien_lattice_2012,murthy_lattice_2018,escande_lattice_2020} and the application to problems involving fractures has been considered in~\cite{chopard_lattice_1998,schluter_lattice_2018,schluter_boundary_2021}.
LBMs are mesoscale methods, employing elements of statistical mechanics, and work with a rather simple transport mechanism on a regular grid. This promises good computational efficiency, especially when the algorithm is parallelized~\cite{kuznik_lbm_2010,mora_concise_2020}.
This work regards the dynamics of antiplane shear, which reduces the Navier-Cauchy equation to a 2D wave equation. Different LBMs~\cite{chopard_lattice_1998,yan_lattice_2000,frantziskonis_lattice_2011} have been proposed to solve the problem of wave propagation. Chopard and Luthi \cite{chopard_lattice_1999} even proposed an LBM to simulate crack growth in a simplified solid. However, this model is not consistent with linear elastic solid mechanics, nor does it include a fracture criterion based on theories of classical fracture mechanics.

In a previous work~\cite{schluter_lattice_2018}, we have shown the LBM for antiplane shear, based on the formulation of Yan's~\cite{yan_lattice_2000} LBM for waves, and applied this to a stationary crack with mode III opening.
We then improved thereon with the introduction of non--lattice conforming boundary conditions~\cite{schluter_lattice_2022}, which are highly relevant for the utilization of the LBM in the field of fracture mechanics.

Now we directly build on this by introducing the extension to dynamic crack propagation.
It is our intention to demonstrate that crack growth can be modeled with the LBM in a way that is in agreement with classical fracture mechanics. However, the comparison to competing approaches  such as finite element simulations of phase field models for fracture~\cite{kuhn_continuum_2010, borden_phase-field_2012} and the {X-FEM}~\cite{moes_finite_1999} is beyond the scope of this work.
In essence, handling dynamic cracks is reduced to a post-processing of lattice sites, without a need for remeshing due to the fixed grid. The criterion for crack propagation that we employ is based on the concept of stress intensity factors (SIFs) by Irwin~\cite{irwin_analysis_1957}. The SIF characterizes the load on the crack tip and is evaluated via the elastic fields in the vicinity of the crack tip in our approach.

Our work is structured as follows.
First a short background on continuum and fracture mechanics is given in Sec.~\ref{sec:mechanics}, followed by a review of the LBM for waves in Sec.~\ref{sec:lbm}.
Next, in Sec.~\ref{sec:implementation}, we present the algorithm and discuss details regarding the implementation of crack propagation and fracture criteria.
In Sec.~\ref{sec:numerics} numerical results are shown for two problems. In order to demonstrate the agreement with classical fracture mechanics and to increase the reliability of our approach, we deliberately chose relatively simple problems for which an analytical benchmark solution exists.
In the first example we do not employ any fracture criterion to decide whether a crack will propagate, but simply force the crack to grow at a certain speed. For this case, we demonstrate that the LBM is able to predict the elastic fields surrounding a moving crack tip, i.e. the SIF, in agreement with analytic solutions.

  In the second numerical example, the implementation of the fracture criterion is tested and it is shown that, in our algorithm, a crack propagates if the fracture criterion is fulfilled.

Lastly, Sec.~\ref{sec:conclusion} summarizes the results. It also discusses the algorithm and its generalization to different LB schemes, such as plane strain, or different propagation criteria.

\section{Mechanics of Linear Elastic Solids and Fractures}
\label{sec:mechanics}

This section summarizes the concepts of solid mechanics and fracture mechanics that the proposed new LBM builds upon.

\subsection{Antiplane Shear Deformation of Linear Elastic Solids}
\label{sec:mechanics-solid}

\begin{figure}[bt]
    \centering
    \includegraphics[width=0.5\textwidth,keepaspectratio]{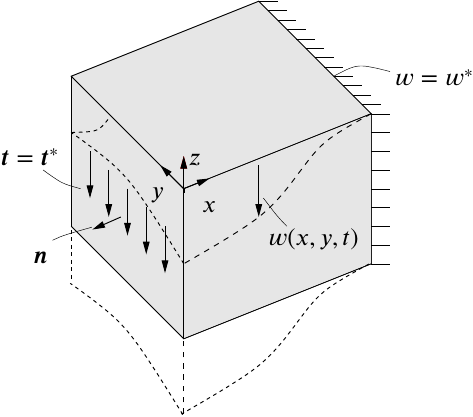}
    \caption{Antiplane shear deformation  of a linear  elastic solid with outer normal vector $\boldsymbol{n}$ subjected to Neumann boundary conditions $\boldsymbol{t}^*$ and Dirichlet boundary conditions $w^*$, cf.~Schl\"uter et al. (2018)~\cite{schluter_lattice_2018}}. The outline of the body in the deformed configuration is indicated by dashed lines.
    \label{fig:antiplane-shear}
\end{figure}

A linear elastic body with shear modulus $\mu$ in a Cartesian $x,y,z$-coordinate system is considered, for which the displacement field is restricted to \mbox{$\boldsymbol{u}=w(x,y)\boldsymbol{e}_z$}. Here, $x / y$ are the in-plane coordinates and $\boldsymbol{e}_z$ is the unit vector in the out-of-plane direction, see Fig.~\ref{fig:antiplane-shear}. This deformation, in which  the out-of-plane displacement is the only nonzero displacement component and is a function of the in-plane coordinates, is commonly referred to as antiplane shear deformation.

For a linear elastic body with density $\rho$, the governing equation of the displacement field is the wave equation
\begin{equation}
    \frac{1}{c_s^2} \dfrac{\partial^2{w}}{\partial{t}^2} = \dfrac{\partial^2{w}}{\partial{x}^2} + \dfrac{\partial^2{w}}{\partial{y}^2},
    \quad \text{where} \; c_s=\sqrt{\dfrac{\mu}{\rho}}.
\end{equation}

\subsection{Dynamic Linear Elastic Fracture Mechanics}
\label{sec:mechanics-crack}

For sufficiently brittle materials, the debonding of the material during fracturing is determined by the mechanical fields directly in the vicinity of the so-called process zone in which material separation actually takes place.
It is commonly assumed that in such a case, the fields surrounding the process zone can still be determined with sufficient accuracy by the theory of linear elasticity. Thus, fracture criteria are constructed by characterizing the elastic fields in the vicinity of crack tips and comparing these to critical values, which need to be determined in experiments for a given crack loading mode and material.

In this work, attention is restricted to one such fracture criterion in order to model crack growth in brittle materials. The elastic fields surrounding the crack tip are obtained by the LBM and are fed into the fracture criterion that eventually determines whether a crack propagates.

\begin{figure}
    \centering
    \includegraphics[width=0.35\textwidth]{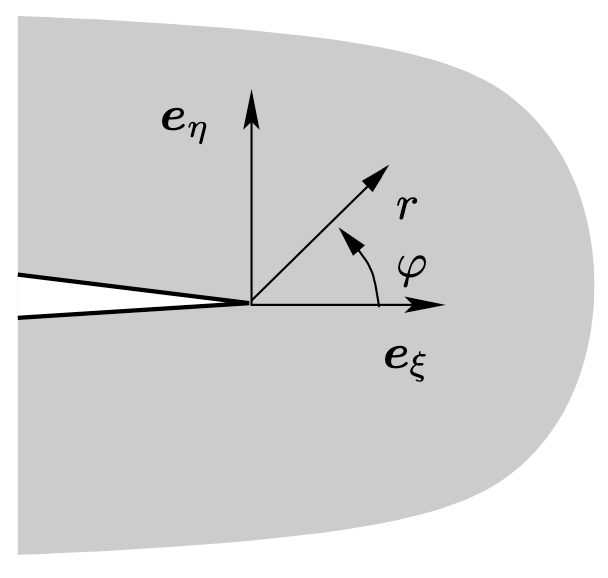}
    \caption{Local crack tip coordinate system. The coordinate $\xi$ represents orientation of the crack tip.}
    \label{fig:crack-tip-cs}
\end{figure}

The criterion employed in this work is Irwin's stress intensity factor (SIF)~\cite{irwin_analysis_1957} criterion, which states that for a given crack deformation, i.e. a particular crack opening mode, the elastic fields in the vicinity of the crack field are a dominated by a universal function in a polar coordinate system attached to the crack tip, see Fig.~\ref{fig:crack-tip-cs}, in which the only unknown is a single scalar parameter. This parameter determines the `intensity' of the loading in the process zone and is referred to as the SIF $K$. For pure antiplane shear loading the displacement field in the vicinity of the crack tip is approximated in terms of $K$ by
\begin{equation}
    w(r,\varphi) \approx \dfrac{2K}{\mu}\sqrt{\dfrac{r}{2\pi}}\sin\left(\frac{\varphi}{2}\right).
    \label{eq:nearTipDisplacement}
\end{equation}
Equation (\ref{eq:nearTipDisplacement}) can be used to determine $K$ for a given, i.e. simulated, displacement field. For example, in a problem with a characteristic length scale $L$ the displacement jump across a crack at a small distance $r\ll{L}$ away from the crack is given by
\begin{equation}
    \delta = \abs{w(r,\pi) - w (r,-\pi)} = \dfrac{4K}{\mu}\sqrt{\dfrac{r}{2\pi}},
\end{equation}
which can be solved for $K$ as
\begin{equation}
   K = \delta\dfrac{\mu}{4}\sqrt{\dfrac{2\pi}{r}}, \quad \text{for } r\ll{L}.
   \label{eq:sif}
\end{equation}
The current stress intensity can subsequently be compared to a material specific critical SIF $K_{C}$ at which crack growth occurs in experiments in order to obtain a crack growth criterion given by
\begin{equation}
    K=K_C,
\label{eq:sifCriterion}
\end{equation}
i.e. a crack will grow if the SIF reaches the critical value.

\section{Lattice Boltzmann Method for the Wave Equation}
\label{sec:lbm}

\begin{figure}[tbp]
    \centering
        \includegraphics[keepaspectratio,width=0.6\textwidth]{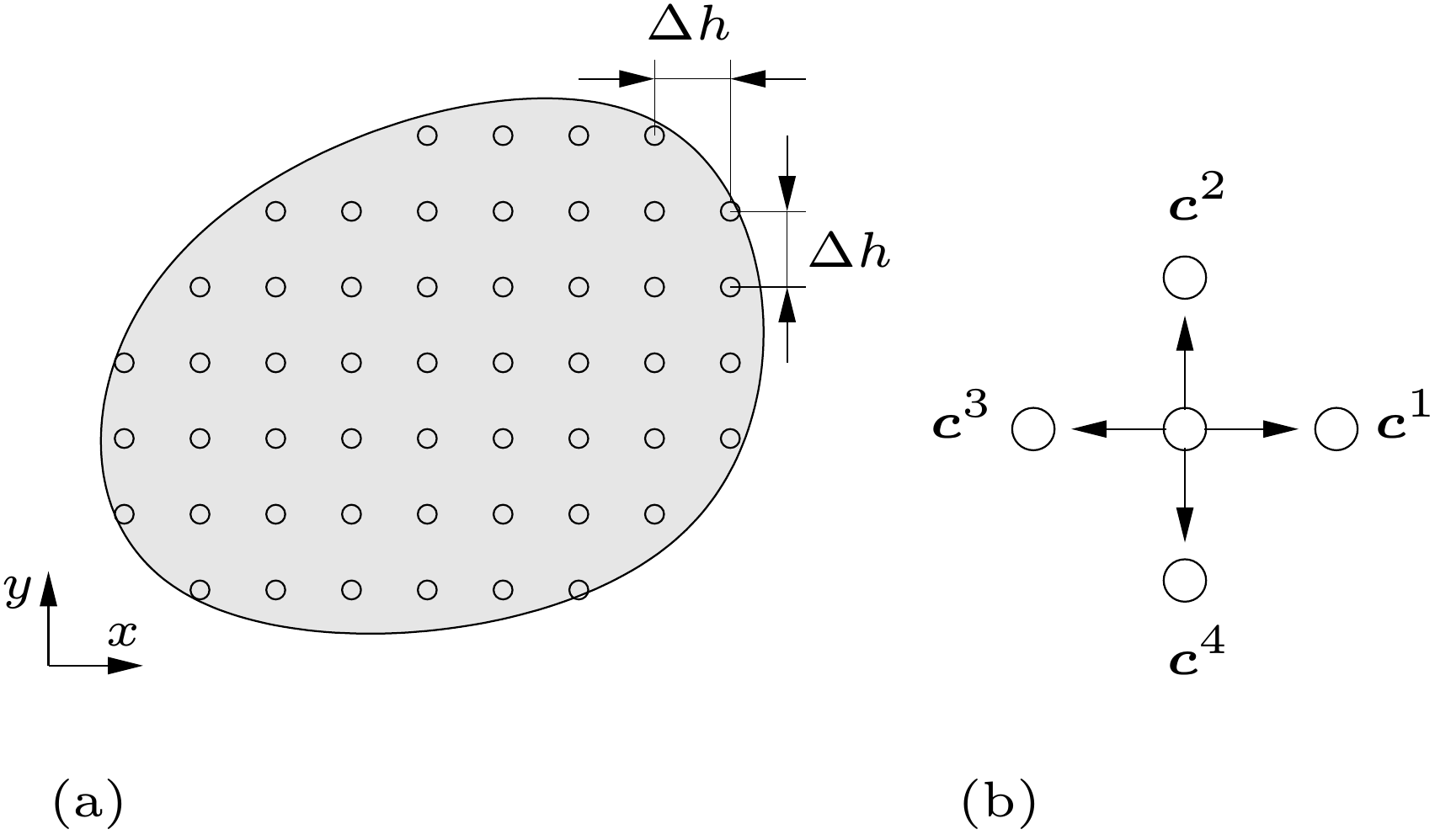}
    \caption{(a) Lattice representation $\mathfrak{B}$ of the elastic solid and
        (b) and the associated lattice velocity vectors (lattice links) for a single lattice point, cf.~Schl\"uter et al. (2018)~\cite{schluter_lattice_2018}.
    }
    \label{fig:domain}
\end{figure}

The LB scheme for wave equations, as proposed by Yan~\cite{yan_lattice_2000}, was already presented in~\cite{schluter_lattice_2018} and summarized in~\cite{schluter_lattice_2022} as well. Therefore only a short overview is given here, which aims at readers that are already familiar with the LBM in general.

The body $\mathcal{B}$, is discretized by a lattice representation $\mathfrak{B}$ with uniform spacing $\Delta h$, see Fig.~\ref{fig:domain}.
A D2Q5 lattice scheme is used, meaning that five  distribution functions $f^\alpha$ per lattice site are regarded, one for each lattice velocity ${\vec{c}^{\alpha}}$, ${\alpha \in \left\{ 0,1,2,3,4 \right\}}$, defined as
\begin{gather}
    \vec{c}^{0} = \begin{pmatrix} 0 \\ 0 \end{pmatrix}, \quad
    \vec{c}^{1} = \begin{pmatrix} c \\ 0 \end{pmatrix}, \quad
    \vec{c}^{2} = \begin{pmatrix} 0 \\ c \end{pmatrix}, \quad
    \vec{c}^{3} = \begin{pmatrix} \text{-}c \\ 0 \end{pmatrix}, \quad
    \vec{c}^{4} = \begin{pmatrix} 0 \\ \text{-}c \end{pmatrix},
\end{gather}
where $c = \nicefrac{\Delta h}{\Delta t}$ is the lattice speed with a time step $\Delta t$ and $c$ is closely related to the shear wave speed $c_s$.

The lattice Boltzmann equation with the BGKW--collision operator~\cite{bhatnagar_model_1954,welander_temperature_1954} is given by
\begin{equation}
    f^{\alpha} \left( \vec{x} + \vec{c}^{\alpha} \Delta{t}, t + \Delta{t} \right) = f^{\alpha} \argum - \frac{\Delta t}{\tau} \Big[ f^{\alpha} \argum - f^{\alpha}_{\text{eq}} \argum \Big].
    \label{eq:LBE}
\end{equation}

For the wave equation, the distribution functions are connected to the particle velocity
\begin{gather}
    \frac{\partial w \argum}{\partial t} = \dot{w} \argum = \sum_\alpha f^{\alpha} \argum
    \label{eq:distrib-macro}
\end{gather}
and the equilibrium distributions for the wave equation are defined as
\begin{gather}
    \begin{aligned}
        f_\text{eq}^0 &= \frac{\partial w}{\partial t} - \frac{2 \lambda w}{c^2},   \\
        f_\text{eq}^{\kappa} &= \frac{\lambda w}{c^2} \quad \text{for } \kappa \in \{ 1,2,3,4 \},
        \label{eq:f-eq}
    \end{aligned}
    \intertext{where}
    \lambda = \frac{1}{\Delta t \, (\tau - \nicefrac{1}{2})} \quad \text{and} \quad \tau = \Delta t.
    \nonumber
\end{gather}

Finally, the displacement $w$ is computed from the particle velocity $\dot{w}$ by an Euler integration scheme via
\begin{gather}
    w \argum = w (\vec{x}, t - \Delta t) + \Delta t \; \dot{w} \argum.
    \label{eq:integration}
\end{gather}

\subsection{Non--Lattice Conforming Boundary Conditions}
\label{sec:lbm-bc}

\begin{figure}[tbp]
    \centering
    \includegraphics[]{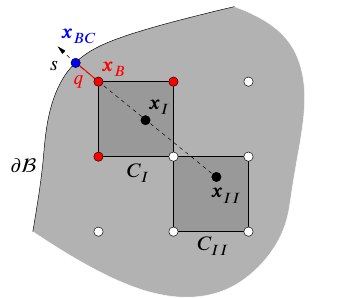}
    \caption{A section of the considered body where non--lattice conforming boundary conditions are applied at boundary lattice point $\boldsymbol{x}_B$.}
    \label{macr_setup}
\end{figure}

Boundary conditions, which are able to accommodate arbitrary boundary geometries, are important for the accurate representation of a crack within the body~$\mathcal{B}$. For the LBM used in this work, two strategies have been proposed in~\cite{schluter_lattice_2022}.
A summary of the macroscopic strategy is given here, since boundary handling is relevant to the algorithm proposed in Sec.~\ref{sec:algo}.

The LB equation~(\ref{eq:LBE}) is able to determine the evolution of the distribution functions, and thus also the evolution of the macroscopic displacement, in the interior of the domain. However, lattice points at the boundary miss one or more lattice links, i.e. neighbor lattice points. These lattice points are denoted as boundary lattice points $\boldsymbol{x}_B$. For boundary lattice points, not all distribution functions can be obtained from collision and streaming, i.e. by the lattice Boltzmann equation~(\ref{eq:LBE}). Instead, the missing distribution functions have to be determined by the boundary conditions.

In this work, we employ a macroscopic algorithm that is capable of handling non--lattice conforming boundary conditions. The algorithm approximates the displacement field $w$ in the vicinity of a particular boundary lattice point by means of a quadratic polynomial that is determined by the value of the boundary condition as well as the displacement field at two interior points $\boldsymbol{x}_I$ and $\boldsymbol{x}_{II}$, see Fig.~\ref{macr_setup}. The value of the displacement field at the interior points is determined by bilinear interpolation inside the cells $C_I$ and $C_{II}$. Evaluating the resulting polynomial at $\boldsymbol{x}_B$ yields a linear equation in terms of the unknown displacements at boundary lattice points for each $\boldsymbol{x}_B$. These equations are assembled in a linear system of equations of the form
\begin{gather}
    \boldsymbol{S}(t+\Delta{t})\boldsymbol{w}_B(t+\Delta{t}) = \boldsymbol{R}(t+\Delta{t}),
\end{gather}
where $\boldsymbol{S}$ only contains information depending on discretization and geometry, and  $\boldsymbol{R}$ also involves the current value of the boundary conditions. Note that $\boldsymbol{S}$ is time dependent if geometry and lattice change over time as is the case during crack propagation.

The system of equation is subsequently solved for the unknown displacements at the boundary lattice points $\boldsymbol{w}_B$, which is then used to determine the value of the missing distribution functions according to
\begin{align}
\begin{split}
    & f^{\alpha}(\boldsymbol{x}_B,t+\Delta{t})  = \dfrac{1}{n_\text{miss}}\left[ \dfrac{w(\boldsymbol{x}_B, t+\Delta{t}) - w(\boldsymbol{x}_B, t)}{\Delta{t}}-\sum_{\beta\in{\cal F}_{\boldsymbol{x}_B}}f^{\beta}(\boldsymbol{x}_{B},t+\Delta{t})\right], \\  & \forall \alpha \notin {\cal F}_{\boldsymbol{x}_B}
    \label{eq:f-macro-bc},
\end{split}
\end{align}
where ${\cal F}_{\boldsymbol{x}_B}$ is the set of  distribution functions that can be determined by the LB equation (\ref{eq:LBE})  and $n_{\text{miss}}$ is the number of missing distribution functions at $\boldsymbol{x}_B$ that cannot.

The part of the algorithm that deals with the implementation of the boundary conditions  determines the macroscopic field $w(\boldsymbol{x}_B,t+\Delta{t})$ at each boundary lattice point such that it is consistent with the boundary conditions on the macroscopic scale. This is why, we refer to the algorithm as a `macroscopic' algorithm for the treatment of the boundary conditions.

\section{Implementation}
\label{sec:implementation}

This section describes the algorithm and considerations for the implementation in general terms. Further details can be found in Sec.~\ref{sec:numerics}, where numerical models and results are discussed.

\subsection{Concepts and Algorithm of Crack Propagation}
\label{sec:algo}

\begin{figure}[bp]
    \centering
    \begin{subfigure}[t]{36mm}
        \centering
        \includegraphics[keepaspectratio,width=\textwidth]{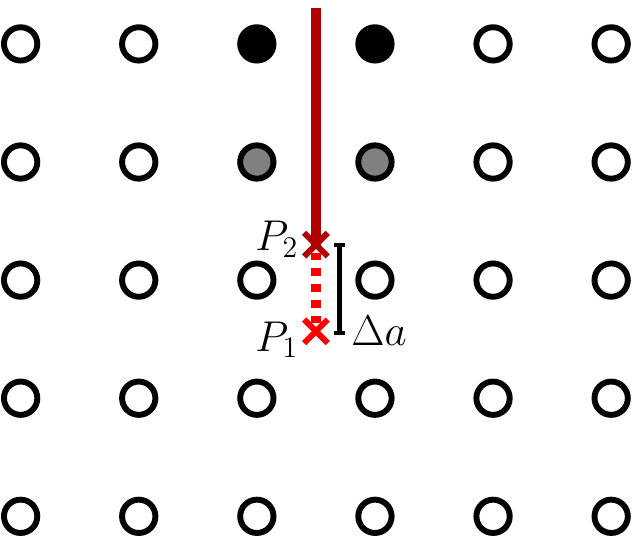}
        \subcaption{The crack is extended by a segment (stippled line) of length $\Delta a = \dot{a} \, \Delta t$, with $P_1$ as the new crack tip.
            }
        \label{fig:crack-growth-1}
    \end{subfigure}
    \hfill
    \begin{subfigure}[t]{36mm}
        \centering
        \includegraphics[keepaspectratio,width=\textwidth]{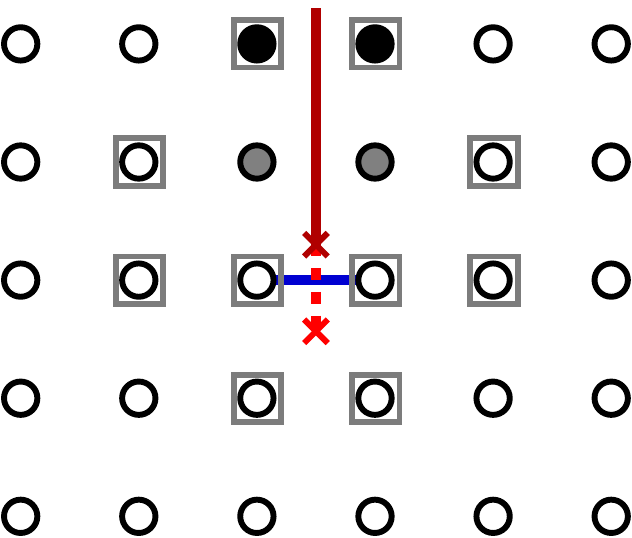}
        \subcaption{For finding the intersected link (blue line), all links connected to points in gray boxes are successively checked.
            }
        \label{fig:crack-growth-2}
    \end{subfigure}
    \hfill
    \begin{subfigure}[t]{36mm}
        \centering
        \includegraphics[keepaspectratio,width=\textwidth]{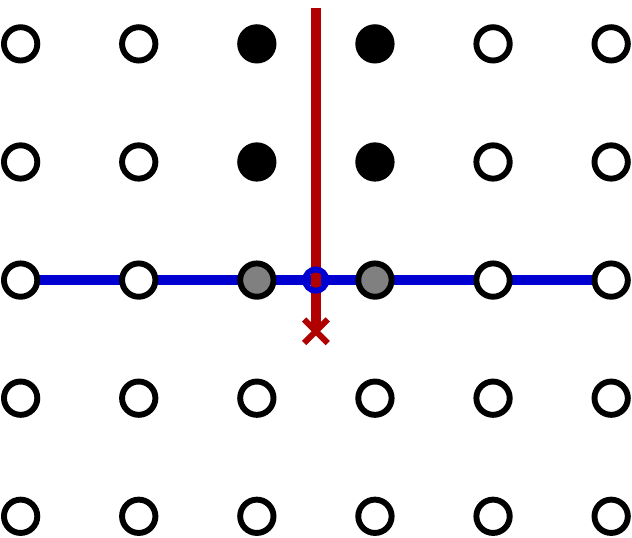}
        \subcaption{The new boundary points are processed. The interpolation involves the points on the blue line, including the closest point on the crack.
            }
        \label{fig:crack-growth-3}
    \end{subfigure}
    \caption{Handling of a dynamic crack (dark red) with processing of boundary points (filled circles).
        }
    \label{fig:crack-growth}
\end{figure}

The propagation of dynamic cracks is handled mostly in a rather simple geometric manner.
An initial crack is needed. It is modeled as a line and independent of the lattice. This also defines the initial crack tip, viewed as a point, and the boundary conditions. Additionally the direction $\hat{c}$ of crack growth is needed. This restricts problems to straight cracks, 
but in turn the lattice can be easily aligned to single cracks.
Our implementation allows two types of simulations. First, it is possible to prescribe crack growth at a constant rate $\dot{a} = \nicefrac{\Delta a}{\Delta t}$. The second option is to let crack growth be determined by the $K$-criterion, based on the SIF. For this case the critical value $K_C$ at which a crack will grow must be specified.

A time step\footnotemark{} of the LBM ends with the time integration~\eqref{eq:integration}, thus the displacement field is fully updated at every point of the lattice~$\mathfrak{B}$.
The handling of crack propagation is then appended to the end as an additional step.
This step itself is subdivided, as described in Alg.~\ref{alg:crack}. First the criterion is evaluated. This is trivial for steady growth. For the $K$-criterion the SIF is computed according to Eq.~\eqref{eq:sif} and compared to the critical value $K_C$, see also Sec.~\ref{sec:mechanics-crack}. This procedure is repeated for multiple crack tips, if needed.

\footnotetext{
    The steps of the LBM, including boundary conditions, are summarized in Alg.~1 of~\cite{schluter_lattice_2022}.
}

\begin{algorithm}[pt]
\caption{Handling of dynamic cracks in the LBM}\label{alg:crack}
\begin{algorithmic}[1]
\Require criterion, direction $\vec{\hat{d}}$, $\dot{a}$ for steady growth, $K_C$ for K-criterion
\Statex
\Procedure{CrackGrowth}{}
    \State propagation $\gets$ \Call{EvaluateCriterion}{}
    \If{propagation = \texttt{True}}
        \State $\vec{v} \gets \dot{a} \, \cdot \, \vec{\hat{d}}$
        \State $\text{cr\_tip}_\text{prev} \gets \text{cr\_tip}$
        \State $\text{cr\_tip} \gets \text{cr\_tip} + \Vec{v} \, \Delta t$
        \State create cr\_segment$(\text{cr\_tip}_\text{prev} \leftrightarrows \text{cr\_tip}$)
            \Comment new crack segment
        \State $B$ $\gets$ \Call{CheckLinks}{cr\_segment}
        \State \Call{ProcessPoints}{$B$}
    \EndIf
\EndProcedure
\Statex
\Function{CheckLinks}{cr\_segment}
    \State let $\mathcal{N}(p)$ be the set of neighbors linked to lattice points $p$
    \State let $(B_\text{prev})$ be the set of boundary points from previous increment
    \State $Q \gets \mathcal{N}(B_\text{prev})$
        \Comment queue
    \State $B$ $\gets \emptyset$
        \Comment new boundary points
    \State $V \gets \emptyset$
        \Comment visited
    \While{$\vert Q \vert > 0$}
        \State let $p \in Q$
        \State $Q \gets Q \backslash \{ p \}$
            \Comment pop $p$ from queue
        \State $V \gets V \cup \{ p \}$
            \Comment mark as visited
        \ForAll{$n_p \in \mathcal{N}(p)$}
            \State create link$(p \leftrightarrows n_p2)$
            \If{cr\_segment intersects link$(p \leftrightarrows n_p)$}
                \State $B \gets B \cup \{p, n_p\}$
                \State $Q \gets Q \cup \mathcal{N}(n_p) \backslash V$
            \EndIf
        \EndFor
    \EndWhile
    \State \textbf{return} $B$
\EndFunction

\end{algorithmic}
\end{algorithm}

If the crack grows, the crack tip is moved along the direction $\hat{c}$. Subsequently, a new segment is created, see Fig.~\ref{fig:crack-growth-1}, between the previous and the new crack tip position, with length
\begin{gather}
    \Delta a = \dot{a} \, \Delta t
\end{gather}
where $v = \nicefrac{\dot{a}}{c_s}$ is the relative speed.
Since this new segment acts as a boundary within the computational domain, the adjoining lattice points need to be processed accordingly.
Generally, these points must be found first. For each point, every associated lattice link is examined, see Fig.~\ref{fig:crack-growth-2}. These links can be treated as lines connecting the point and its respective neighbor and are checked for an intersection with the crack segment. Any intersected link is subsequently severed and no information is exchanged along it in further time steps. Both associated points need to be processed for the boundary conditions and their implementation, as for the initial crack. For the macroscopic implementation proposed in~\cite{schluter_boundary_2021,muller_lattice_2021} and used throughout the numerical models in Sec.~\ref{sec:numerics}, this entails extending and inverting the boundary coefficient matrix and expanding the vector of interpolation coefficients by the newly found boundary points, see Fig.~\ref{fig:crack-growth-3}. This task is computationally expensive and requires more time, the longer the crack grows.

While the growth of the crack and determining severed lattice links is a universally applicable concept, the processing of new boundary points needs to be adapted to different techniques of boundary handling, e.g.\ when a different LBM for solids is employed.

For this LBM, the lattice wave speed~$c$ should surpass the shear wave speed~$c_s$, i.e.\ $c = \kappa \, c_s$, where $\kappa \geqslant 1$. Since ${\dot{a} < c_s}$, it follows that  ${\Delta a < \nicefrac{\Delta h}{\kappa} \leqslant \Delta h}$. Thus
only one pair of new boundary points is expected, at most, for straight cracks. Thus the number of links to be checked can be severely reduced by regarding only the immediate vicinity of the crack tip. A queue is generated from the neighbors of the last pair of boundary points that have been found. For every point in the queue, the associated links need to be checked. Once a severed link is found, the neighbors of the linked points are added to the queue. The number of checks to be performed can reduced further by marking points that have been completely visited, i.e. all associated links have been checked.

This type of processing also works for different kinds of LB methods, e.g. on D2Q9 lattices, and for cases in which the direction of crack propagation is not prescribed.

\subsection{Further Details for the $K$-Criterion}
\label{sec:algo:details}

\subsubsection*{Evaluation of Stress Intensity Factors}

\begin{figure}[tbp]
\begin{subfigure}[t]{0.485\textwidth}
    \centering
    \includegraphics[width=\textwidth]{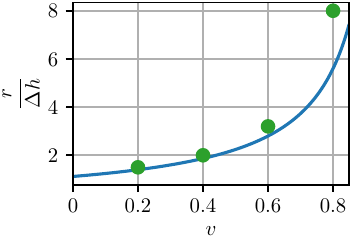}
    \caption{The minimum distance of evaluation $r_\text{min}$ \eqref{eq:r(v)} for $r_0 = 0.07 L$ (blue line). The green dots represent values chosen for steady growth (see Fig.~\ref{fig:steady-results}).}
    \label{fig:r(v)}
\end{subfigure}
\hfill
\begin{subfigure}[t]{0.485\textwidth}
    \centering
    \includegraphics[width=\textwidth]{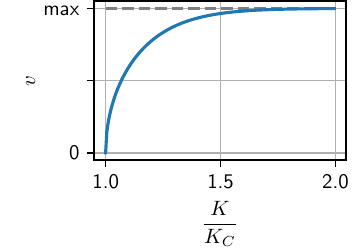}
    \caption{Regularization of the crack velocity $v$ \eqref{eq:v-tanh} with respect to the relative overshoot of the SIF $\nicefrac{K}{K_C}$.}
    \label{fig:tanh}
\end{subfigure}
\caption{Functions introduced to define continuous parameters for the K-criterion.}
\end{figure}

\begin{figure}[tbp]
    \centering
    \includegraphics[width=64mm]{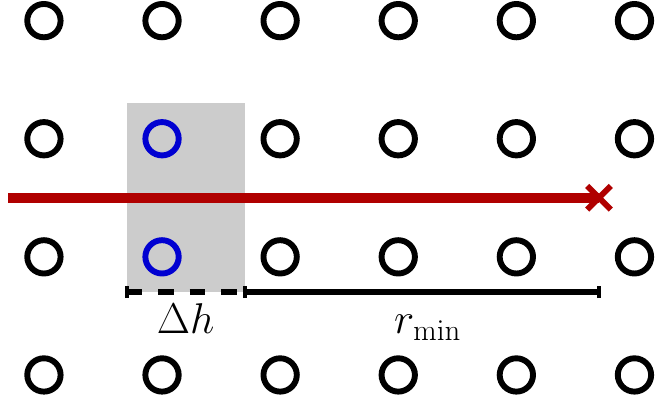}
    \caption{Points (blue circles) along the crack (red line) at a distance ${r \in [r_\text{min}, r_\text{min} + \Delta h]}$ (gray area) are used for the for evaluation of the SIF.
    }
    \label{fig:sif-r}
\end{figure}

The evaluation of the SIF is carried out according to Eq.~\ref{eq:sif}. The crack opening displacement $\delta$ is computed as the difference between $w$ at two lattice points adjacent to, but not on the crack itself, see Fig.~\ref{fig:sif-r}.
For stationary cracks, such as in~\cite{schluter_lattice_2018,schluter_boundary_2021}, this is straight forward, with a clearly defined distance $r$ from the crack tip. The SIF can be determined in a post-processing step.
However, with a growing crack $r$ varies between time steps and for the $K$-criterion especially. In addition, the SIF has to be evaluated in every time step.
The lattice points for evaluation are chosen, such that their distance $r$ to the crack tip lies within an interval $[r_\text{min}, r_\text{min} + \Delta h]$. As reported in~\cite{schluter_lattice_2018,schluter_boundary_2021}, the results for the SIF are closer to analytical values when the evaluation occurs at a distance from the crack tip, which is indicated here by the parameter $r_\text{min}$.
The SIF $K$ is rather sensitive to $r$, thus for the $K$-criterion this parameter should be adaptable during runtime.
Preliminary numerical results showed a correlation with $v$, which can be modeled by
\begin{gather}
    r_\text{min} = \frac{r_0}{1-v}, \quad r_0 \equiv r(v=0).
    \label{eq:r(v)}
\end{gather}
This function is shown in Fig.~\ref{fig:r(v)}, together with the values of $r_\text{min}$ used for steady growth in Sec.~\ref{sec:numerics-steady}.
It is chosen purely from empirical considerations.

\subsubsection*{Regularization of the Crack Velocity}

With the discretization of time, crack propagation is discretized as well, allowing the crack to grow by a finite length $\Delta a$ within $\Delta t$.
When using the $K$-criterion to determine the crack propagation, $K$ is evaluated at the end of each time step and might surpass the critical value $K_C$.
This kind of overshooting the critical SIF should be avoided since in real systems, the crack growth would continue to such a state that $K \leqslant K_c$.

In order to reduce the unphysical overshoot of the SIF, the crack velocity is allowed to increase, up to the maximum crack velocity of $v_\text{max}$. Here, the crack velocity is assumed to be a continuous function of $K$,
\begin{gather}
    v \left( K; K_C \right) \thicksim v_\text{max} \, \tanh \left( \sqrt{\left( \frac{K}{K_C} \right)^4 - 1} \right), \quad K > K_C,
    \label{eq:v-tanh}
\end{gather}
see Fig.~\ref{fig:tanh}.

\section{Numerical Results}
\label{sec:numerics}

This section shows numerical results to validate the algorithm and its implementation presented in the previous section.
The first examples verifies the evaluation of $K$ in a dynamical model. Since this has not been done before with a propagating crack, a problem with an analytical solution is used for steady crack growth.
The second example delivers proof of concept for the $K$-criterion. The numerical results are assessed with regard to plausibility.

\subsection{Steady Crack Growth in a Semi-Infinite Strip}

\begin{figure}[bt]
\begin{subfigure}[t]{0.45\textwidth}
    \centering{}
    \includegraphics[width=\textwidth]{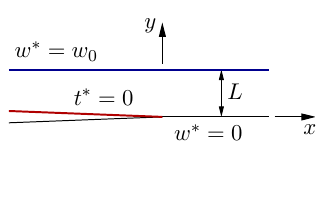}
    \subcaption{}
    \label{fig:mandal-1}
\end{subfigure}
\hfill
\begin{subfigure}[t]{0.45\textwidth}
    \centering{}
    \includegraphics[width=\textwidth]{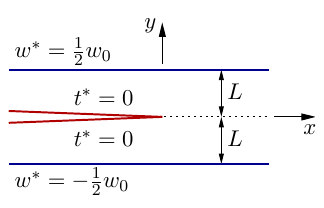}
    \subcaption{}
    \label{fig:mandal-2}
\end{subfigure}

\begin{subfigure}{\textwidth}
    \centering
    \includegraphics[]{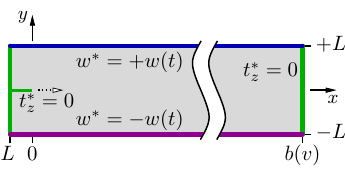}
    \subcaption{}
    \label{fig:strip}
\end{subfigure}

    \caption{(a) Geometry of Mandal's original problem~\cite{mandal_moving_2020} and (b) domain considered for the numerical examination. (c) Numerical realization of this geometry.}
    \label{fig:mandal}
\end{figure}

\subsubsection*{Stress Intensity Factor}
\label{sec:numerics-mandal}

We consider a semi-infinite crack that grows at a constant rate $\dot{a} < c_s$, i.e. $v < 1$, in a semi-infinite elastic body, as shown in Fig.~\ref{fig:mandal}~a). We consider the surface $y=h$ to be subject to a Dirichlet boundary condition
\begin{equation}
    w(x,h) = w_0
\end{equation}
and the (moving) crack faces to be traction-free, i.e.
\begin{equation}
    \sigma_{yz}(\xi<0,\eta=\pm0) = 0.
\end{equation}
In \cite{mandal_moving_2020} it is shown that the steady-state stress intensity factor for the problem described above is
\begin{equation}
    K=-\mu{w_0} \sqrt{\dfrac{2\beta}{L(2\beta{L}+1)}}
    \label{eq:sif-mandal}
\end{equation}
where
\begin{equation}
    \beta = \sqrt{1 - v^2}.
\end{equation}
The original problem depicted in Fig.~\ref{fig:mandal}~a) implies that $w=0$ at the lower crack face. In this work, a slightly different but related problem is simulated, see Fig.~\ref{fig:mandal}~b). We apply
\begin{equation}
    w(x,\pm{L})=\pm \tfrac{1}{2}w_0
\end{equation}
at the top and bottom edges of a strip with width $2L$ in which a traction-free crack propagates at a steady velocity in $x$-direction. Although only half the displacement, i.e. $\frac{1}{2}w_0$, is applied at the top and bottom edge compared to the original analytical solution in which only the top edge is loaded, see Fig.~\ref{fig:mandal}~a), we expect the total crack opening and the stress intensity factor for the problem Fig.~\ref{fig:mandal}~b) to be the same as in the original analytical solution, since both crack faces are displaced. The infinite strip problem from Fig.~\ref{fig:mandal-2} is eventually approximated by a finite domain in our simulations, see Fig.~\ref{fig:strip}.

\begin{table}[t]
    \caption{Numerical and statistical data regarding steady crack growth for different values of $v$; $K^\text{theo}$ is the expected value from \eqref{eq:sif-mandal}. The mean with standard deviation $\sigma$ and median with differences to the 25th and 75th percentile (cf. Fig.~\ref{fig:steady-results}) are gathered from
    experimental values for $K$, evaluated at $\nicefrac{r_\text{min}}{\Delta h}$.
    }
    \centering
    \begin{tabular}{ *{7}{c|} c}
        $v$ & $\nicefrac{r_\text{min}}{\Delta h}$ & $K^\text{theo}$ & mean & $\pm \sigma$ & median & $-25\%$ & $+75\%$   \\
        \hline
        0.2	& 1.50  & 0.1627    & 0.1664	& 0.0083	& 0.1640	& 0.0036	& 0.0064    \\
        0.4	& 2.25	& 0.1609    & 0.1642	& 0.0057	& 0.1624	& 0.0035	& 0.0068    \\
        0.6	& 4.00  & 0.1569    & 0.1576	& 0.0029	& 0.1584	& 0.0033	& 0.0018    \\
        0.8	& 8.00  & 0.1477    & 0.1475    & 0.0007	& 0.1477    & 0.0008	& 0.0005
    \end{tabular}
    \label{tab:steady}
\end{table}

\begin{figure}[t]
    \centering
    \includegraphics[width=0.9\textwidth]{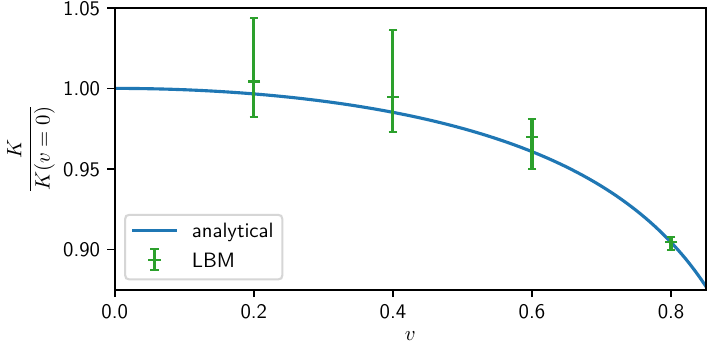}
    \caption{SIFs for different crack velocities $v$ in comparison to the analytical values of Eq.~\eqref{eq:sif-mandal}. The values represent the median with errorbars indicating the range from the 25th to the 75th percentile.}
    \label{fig:steady-results}
\end{figure}

\subsubsection*{Validation}
\label{sec:numerics-steady}

The analytical solution for the SIF by Mandal~\cite{mandal_moving_2020}
is based on the assumption of a quasi-stationary problem. This implies certain constraints  on the geometry  of the domain and the prescribed boundary conditions. As mentioned before, we approximate the infinite strip Fig.~\ref{fig:mandal}~b) by the domain Fig.~\ref{fig:mandal}~c).
The left and right edges, as well as the crack faces, are free boundaries. On the upper and lower edge the displacement is prescribed by
\begin{gather}
    w(t) = \tfrac{1}{2} w_0 \cdot \begin{cases}
        \sin^2 (\tfrac{\pi}{2} \tfrac{t}{t_0}),  & t < t_0, \\
        1,                                              & t \geqslant t_0.
    \end{cases}
\end{gather}
Since sudden changes in the load can excite spurious waves, the displacement is gradually increased to its final value of $\frac{1}{2} w_0$ at the time $t_0$, where $t_0$ is chosen in relation to the rate of crack growth. The crack continues to grow in a quasi-stationary period until a time of $t_f > t_0 + 15 \, \nicefrac{c_s}{L}$.

The initial crack is rather short compared to $r_\text{min}$, with a length of $0.5 L$.
But it grows during the start up period and can be considered to be semi-infinite between $t_0$ and $t_f$. The position $b$ of the right edge is chosen, such that it does not influence the fields around the crack tip at $t_f$, i.e. $b(v) > v c_s \, t_f + 2 L$. Thus both the start up time $t_0$ and the total length of the strip have to increase with the relative speed~$v$ in order to yield comparable results to the infinite strip.
This numerical experiment has been conducted for different crack speeds, i.e.  ${v = 0.2, 0.4, 0.6}$~and~$0.8$, with a lattice spacing of ${\Delta h = \nicefrac{L}{16}}$.

As described in Sec.~\ref{sec:algo:details}, $r_\text{min}$ is adjusted for each value of $v$.
But the actual value of $r$ differs between time steps within a certain interval, cf.~Fig.~\ref{fig:sif-r}, because the crack tip position changes relative to the lattice points that are used to evaluate $K$.
Due to the sensitivity of $K$ regarding $r$, this causes $K$ to fluctuate around the expected steady value.
Thus a statistical evaluation is undertaken with a total of $300$~data points, that are sampled for $t \in [t_f - 15 \, \nicefrac{c_s}{L}, t_f]$.
From this data, the arithmetic mean value with the standard deviation and the median value with the 25th- and 75th-percentile are gathered.
This evaluation is compiled in Tab.~\ref{tab:steady} and the median is also depicted in Fig.~\ref{fig:steady-results}.
The results are close to the analytical curve, each within the margin of error
indicated by the percentiles.
This margin gets smaller for higher $v$.
The values of $K$ lie above the analytical values, with the exception of $v=0.8$, where it is easier to adjust $r_\text{min}$.

\begin{figure}[t]
    \centering
    \includegraphics[]{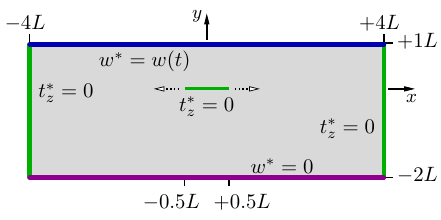}
    \caption{Domain for the validation of the K-criterion}
    \label{fig:k-crit-domain}
\end{figure}

\subsection{Validation of the K-criterion}
\label{sec:numerics:krit}

The domain for this example is a rectangle of size $3L$ by $8L$ with an initial crack of length $1L$ along the $x$-axis, within $x=\pm 0.5L$. The crack is located at $1L$ from the upper and $2L$ from the lower edge, as depicted in Fig.~\ref{fig:k-crit-domain}. The lattice spacing is given by $\Delta h = 2^{-6} \, L$, resulting in a total of
$98\,304$ lattice points.

Both crack tips can propagate horizontally along the $x$-axis.
For this example, ${K_C = 0.0055 \, \mu \sqrt{L}}$ is chosen, with ${r_0 =0.03L \approx 2 \, \Delta h}$ for $r_\text{min}$ in Eq.~\eqref{eq:r(v)}
and ${v_\text{max} = 0.85}$ as the maximum velocity.
For all boundaries the macroscopic implementation of non--lattice conforming boundary conditions, as described in \cite{schluter_lattice_2022}, is used.
The lower edge is subjected to homogeneous Dirichlet boundary conditions $w^* = 0$. Both lateral edges and the crack faces are traction-free with $t^*_z = 0$, while the upper edge has a time-dependent Dirichlet boundary condition prescribed by
\begin{gather}
    w(t) = 0.01 L \, \begin{cases}
    \sin (\tfrac{\pi}{8} t), & t < 8 \, \tfrac{L}{c_s}, \\
    0,                              & t \geqslant 8 \, \tfrac{L}{c_s},
    \end{cases}
\end{gather}
which is a half-period of a sine-function.
This excites an elastic wave, which then propagates through the domain and is reflected at the outer edges and the crack.

\begin{figure}[tb]
    \centering
    \includegraphics[width=\textwidth]{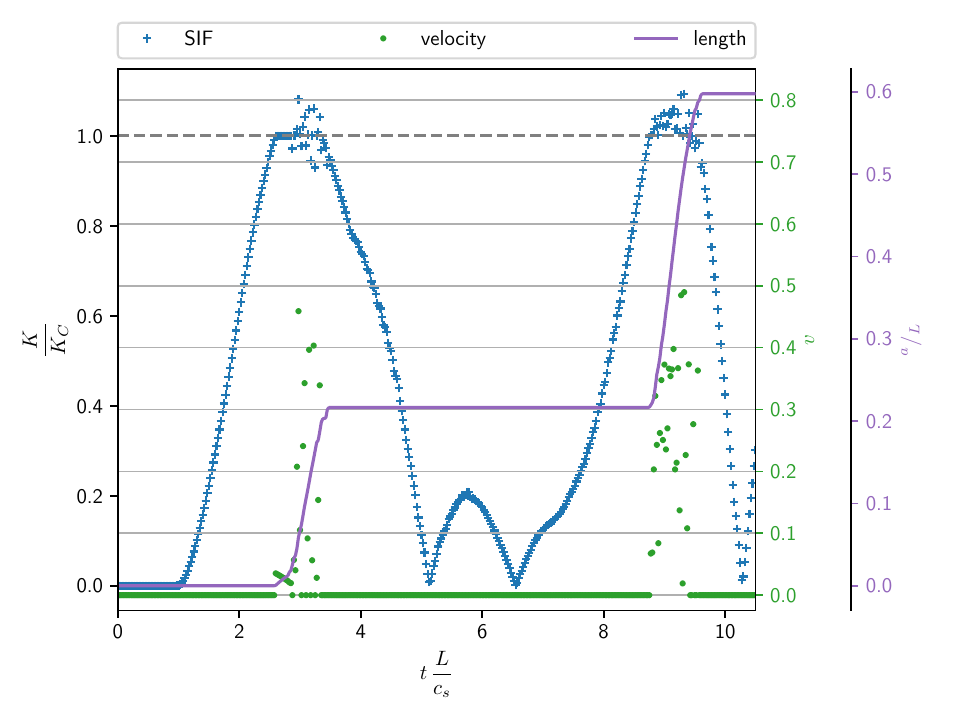}
    \caption{Stress intensity factors, velocity of crack propagation and total additional length for the example of crack growth with the K-criterion, with a maximum allowed velocity $v_\text{max} = 0.85$.
    Due to symmetry, only data for the crack tip propagating in positive $x$-direction is shown}
    \label{fig:k-crit-results}
\end{figure}

Upon reaching the crack, the incident waves cause the SIF to rise, as can be seen in Fig.~\ref{fig:k-crit-results}.
When $K_C$ is surpassed at a crack tip, the increment $\Delta a = v \, \Delta t$ is computed by means of the function $v(K; K_C)$ as defined in Eq.~\eqref{eq:v-tanh}.
Fig.~\ref{fig:k-crit-results} shows the SIF before the crack grows, thus $K$ can be still be higher than $K_C$, in spite of the modification~\eqref{eq:v-tanh}. In the next iteration, $K$ should be close to $K_C$. However, since the configuration changes dynamically, $K$ can potentially surpass the critical value again.

The initial wave leads to an increase in $K$, such that both crack tips initiate crack growth.
The subsequent reflected wave also leads to a propagation of the crack tips. This time, $K$ exceeds $K_C$ slightly more, due to the changed geometry of the domain, and this results in a higher velocity.
During a period of crack growth, $K$ stays at a value close to $K_C$. After the wave has passed, crack growth halts since $K$ decreases again.

\subsection{Remarks on Efficiency}
\label{sec:numerics:efficiency}

To get an impression of the impact the additional crack propagation step has on the efficiency of the LBM, the added computational cost has been examined\footnotemark{}.
For this, the steady growth example with $v=0.4$ has been repeated. $2\,000$ time steps were computed with for a total of $10\,496$ lattice points.
The CPU time spent solely on the crack propagation, but also on the computation in total, was measured.
By design of the example, crack growth occurs in every time step, but only accounts for about $2.8 \%$ of the CPU time.

The example with the criterion was assessed in a similar manner. It ran for a total of $5\,050$ time steps. Here, $0.7 \%$ percent of the CPU time was spent on crack propagation, since the crack only grows in a limited number of time steps, thus effectively skipping the algorithm.

\footnotetext{
    Intel Core I7-1165G7 @ $2.8 \text{GHz} \times 8$; Python 3.9.13
}

\section{Discussion and Conclusion}
\label{sec:conclusion}

This work introduced a method to simulate dynamic crack propagation using an LBM for solids. In contrast to the more established alternatives, such as finite element methods, it is based on the rather efficient concept of processing very few lattice points in each time step. This keeps the regular lattice unchanged, without the need to re-mesh the domain.
While no intensive study on the efficiency has been undertaken so far, it can be projected that the overall efficiency of the LBM is maintained.
Only a small number of lattice points is processed for the identification of new boundary points.
This processing step is only needed if the crack propagates.
Furthermore, boundary conditions are not initialized in each time step that crack growth occurs.
Thus little computational effort is added to the LBM due to this crack propagation step.
This is corroborated by the comparison of CPU times for the examples in Sec.~\ref{sec:numerics:efficiency}.
Additionally, the matrix inversion necessary for the boundary conditions is a computationally expensive operation, which accounts for some of the needed CPU time.
With different boundary conditions, the additional cost could be reduced.

Two cases have been implemented and validated in numerical experiments. In the first one, steady growth of a mode III crack has been compared to analytical results.
This shows a good accuracy for the evaluation of the SIF in a dynamical model, as described in Sec.~\ref{sec:algo:details}.
In the other case, crack propagation with a criterion based on the SIF has been simulated.
This example shows the expected results and delivers evidence, that the algorithm is conceptually capable of describing crack growth based on a fracture criterion.
Difficulties stem from the evaluation of $K$. It is very sensitive to changes in the fields surrounding the crack tip. However, the algorithm introduced here is not tied to the evaluation of $K$.
In fact, not only the method to obtain $K$, but the criterion, as well as the entire LB scheme, could be exchanged and the underlying geometric considerations would still hold.

Moreover, to clarify the generalization beyond the reduced problem of antiplane shear deformation, it can be noted that the algorithm is modular by design.
Effectively, it only depends on the geometric aspects of the lattice.
Since crack propagation is appended as a post-processing step, the LB scheme can be swapped out, as long as it supplies the data necessary for the evaluation of the criterion.
Furthermore, the criterion can be exchanged, e.g.\ to one regarding a mixed mode propagation based on $K_I$ and $K_{II}$ in plane strain.

As topics of future research, this generalization should be undertaken. More studies on the physicality of the results and comparisons to the established numerical methods, especially regarding the computational efficiency, would be of great interest as well.

\section*{Declarations}

\subsubsection*{Funding}
The authors gratefully acknowledge the funding by the German Research Foundation (DFG) within the project~423809639.

\subsubsection*{Code availability}
For access to the git repository please contact the authors.

\clearpage
\bibliographystyle{acm}
\bibliography{SolidLBM}%

\begin{thebibliography}{10}

\bibitem{barsoum_use_1976}
{\sc Barsoum, R.~S.}
\newblock On the use of isoparametric finite elements in linear fracture
  mechanics.
\newblock {\em International Journal for Numerical Methods in Engineering 10},
  1 (1976), 25--37.

\bibitem{bhatnagar_model_1954}
{\sc Bhatnagar, P.~L., Gross, E.~P., and Krook, M.}
\newblock A {Model} for {Collision} {Processes} in {Gases}. {I}. {Small}
  {Amplitude} {Processes} in {Charged} and {Neutral} {One}-{Component}
  {Systems}.
\newblock {\em Physical Review 94}, 3 (May 1954), 511--525.

\bibitem{borden_phase-field_2012}
{\sc Borden, M.~J., Verhoosel, C.~V., Scott, M.~A., Hughes, T.~J., and Landis,
  C.~M.}
\newblock A phase-field description of dynamic brittle fracture.
\newblock {\em Computer Methods in Applied Mechanics and Engineering 217\/}
  (2012), 77--95.
\newblock Publisher: Elsevier.

\bibitem{bourdin_variational_2008}
{\sc Bourdin, B., Francfort, G.~A., and Marigo, J.-J.}
\newblock The {Variational} {Approach} to {Fracture}.
\newblock {\em Journal of Elasticity 91}, 1-3 (Apr. 2008), 5--148.

\bibitem{chopard_lattice_1998}
{\sc Chopard, B., Luthi, P., and Marconi, S.}
\newblock A {Lattice} {Boltzmann} {Model} for {Wave} and {Fracture} phenomena.
\newblock {\em arXiv e-prints\/} (Dec. 1998), cond--mat/9812220.

\bibitem{chopard_lattice_1999}
{\sc Chopard, B., and Luthi, P.~O.}
\newblock Lattice {Boltzmann} computations and applications to physics.
\newblock {\em Theoretical Computer Science 217}, 1 (Mar. 1999), 115--130.

\bibitem{erdogan_numerical_1973}
{\sc Erdogan, F., Gupta, G.~D., and Cook, T.~S.}
\newblock Numerical solution of singular integral equations.
\newblock In {\em Methods of analysis and solutions of crack problems: {Recent}
  developments in fracture mechanics {Theory} and methods of solving crack
  problems}, G.~C. Sih, Ed., Mechanics of fracture. Springer Netherlands,
  Dordrecht, 1973, pp.~368--425.

\bibitem{escande_lattice_2020}
{\sc Escande, M., Kolluru, P.~K., Cléon, L.~M., and Sagaut, P.}
\newblock Lattice {Boltzmann} {Method} for wave propagation in elastic solids
  with a regular lattice: {Theoretical} analysis and validation.
\newblock {\em arXiv:2009.06404 [physics]\/} (Sept. 2020).

\bibitem{frantziskonis_lattice_2011}
{\sc Frantziskonis, G.~N.}
\newblock Lattice {Boltzmann} method for multimode wave propagation in
  viscoelastic media and in elastic solids.
\newblock {\em Physical Review E 83}, 6 (June 2011), 066703.

\bibitem{ha_characteristics_2011}
{\sc Ha, Y.~D., and Bobaru, F.}
\newblock Characteristics of dynamic brittle fracture captured with
  peridynamics.
\newblock {\em Engineering Fracture Mechanics 78}, 6 (Apr. 2011), 1156--1168.

\bibitem{irwin_analysis_1957}
{\sc Irwin, G.~R.}
\newblock Analysis of {Stresses} and {Strains} {Near} the {End} of a {Crack}
  {Traversing} a {Plate}.
\newblock {\em Journal of Applied Mechanics 24}, 3 (Sept. 1957), 361--364.

\bibitem{kruger_lattice_2017}
{\sc Krüger, T., Kusumaatmaja, H., Kuzmin, A., Shardt, O., Silva, G., and
  Viggen, E.~M.}
\newblock {\em The {Lattice} {Boltzmann} {Method}: {Principles} and
  {Practice}}.
\newblock Graduate {Texts} in {Physics}. Springer International Publishing,
  Cham, 2017.

\bibitem{kuhn_continuum_2010}
{\sc Kuhn, C., and Müller, R.}
\newblock A continuum phase field model for fracture.
\newblock {\em Engineering Fracture Mechanics 77}, 18 (2010), 3625--3634.
\newblock Publisher: Elsevier.

\bibitem{kuznik_lbm_2010}
{\sc Kuznik, F., Obrecht, C., Rusaouen, G., and Roux, J.-J.}
\newblock {LBM} based flow simulation using {GPU} computing processor.
\newblock {\em Computers \& Mathematics with Applications 59}, 7 (Apr. 2010),
  2380--2392.

\bibitem{mandal_moving_2020}
{\sc Mandal, P.}
\newblock Moving semi-infinite mode-{III} crack inside the semi-infinite
  elastic media.
\newblock {\em Journal of Theoretical and Applied Mechanics 58}, 3 (July 2020),
  649--659.

\bibitem{marconi_lattice_2003}
{\sc Marconi, S., and Chopard, B.}
\newblock A {Lattice} {Boltzmann} {Method} for a {Solid} {Body}.
\newblock {\em International Journal of Modern Physics B 17}, 01n02 (Jan.
  2003), 153--156.

\bibitem{mora_concise_2020}
{\sc Mora, P., Morra, G., and Yuen, D.~A.}
\newblock A concise python implementation of the lattice {Boltzmann} method on
  {HPC} for geo-fluid flow.
\newblock {\em Geophysical Journal International 220}, 1 (Jan. 2020), 682--702.

\bibitem{moes_finite_1999}
{\sc Moës, N., Dolbow, J., and Belytschko, T.}
\newblock A finite element method for crack growth without remeshing.
\newblock {\em International Journal for Numerical Methods in Engineering 46},
  1 (1999), 131--150.

\bibitem{murthy_lattice_2018}
{\sc Murthy, J., Kolluru, P.~K., Kumaran, V., Ansumali, S., and Narayana~Surya,
  J.}
\newblock Lattice {Boltzmann} {Method} for {Wave} {Propagation} in {Elastic}
  {Solids}.
\newblock {\em Communications in Computational Physics 23}, 4 (2018).

\bibitem{muller_lattice_2021}
{\sc Müller, H., Schlüter, A., and Müller, R.}
\newblock Lattice {Boltzmann} {Method} for {Antiplane} {Shear} with
  {Non}‐{Mesh} {Conforming} {Boundary} {Conditions}.
\newblock {\em PAMM 21\/} (Dec. 2021).

\bibitem{obrien_lattice_2012}
{\sc O'Brien, G.~S., Nissen-Meyer, T., and Bean, C.~J.}
\newblock A {Lattice} {Boltzmann} {Method} for {Elastic} {Wave} {Propagation}
  in a {Poisson} {Solid}.
\newblock {\em Bulletin of the Seismological Society of America 102}, 3 (June
  2012), 1224--1234.

\bibitem{schluter_lattice_2018}
{\sc Schlüter, A., Kuhn, C., and Müller, R.}
\newblock Lattice {Boltzmann} simulation of antiplane shear loading of a
  stationary crack.
\newblock {\em Computational Mechanics 62}, 5 (Nov. 2018), 1059--1069.

\bibitem{schluter_boundary_2021}
{\sc Schlüter, A., Müller, H., and Müller, R.}
\newblock Boundary {Conditions} in a {Lattice} {Boltzmann} {Method} {For}
  {Plane} {Strain} {Problems}.
\newblock {\em PAMM 21\/} (Dec. 2021).

\bibitem{schluter_lattice_2022}
{\sc Schlüter, A., Müller, H., and Müller, R.}
\newblock Lattice {Boltzmann} method for antiplane shear deformation:
  non-lattice-conforming boundary conditions.
\newblock {\em Archive of Applied Mechanics\/} (Aug. 2022).

\bibitem{schluter_phase_2014}
{\sc Schlüter, A., Willenbücher, A., Kuhn, C., and Müller, R.}
\newblock Phase field approximation of dynamic brittle fracture.
\newblock {\em Computational Mechanics 54}, 5 (Nov. 2014), 1141--1161.

\bibitem{silling_meshfree_2005}
{\sc Silling, S., and Askari, E.}
\newblock A meshfree method based on the peridynamic model of solid mechanics.
\newblock {\em Computers \& Structures 83}, 17-18 (June 2005), 1526--1535.

\bibitem{solorzano_lattice_2018}
{\sc Solórzano, S., Mendoza, M., Succi, S., and Herrmann, H.~J.}
\newblock Lattice {Wigner} equation.
\newblock {\em Physical Review E 97}, 1 (Jan. 2018), 013308.

\bibitem{succi_lattice_2018}
{\sc Succi, S.}
\newblock {\em The {Lattice} {Boltzmann} equation: for complex states of
  flowing matter}, first edition~ed.
\newblock Oxford University Press, Oxford, 2018.

\bibitem{welander_temperature_1954}
{\sc Welander, P.}
\newblock On the temperature jump in a rarefied gas.
\newblock {\em Arkiv fysik 7\/} (1954).

\bibitem{yan_lattice_2000}
{\sc Yan, G.}
\newblock A {Lattice} {Boltzmann} {Equation} for {Waves}.
\newblock {\em Journal of Computational Physics 161}, 1 (June 2000), 61--69.

\bibitem{zhang_new_1989}
{\sc Zhang, C., and Achenbach, J.~D.}
\newblock A {New} {Boundary} {Integral} {Equation} {Formulation} for
  {Elastodynamic} and {Elastostatic} {Crack} {Analysis}.
\newblock {\em Journal of Applied Mechanics 56}, 2 (June 1989), 284--290.

\end{thebibliography}

\end{document}